\documentstyle[secnumtab]{fbssuppl}
\title{The Nijmegen Potentials}
\author{J.J. de Swart\thanks{{\it e-mail address:} swart{\tt @}sci.kun.nl},
  R.A.M.M. Klomp, M.C.M. Rentmeester, Th.A. Rijken}
\institute{Institute for Theoretical Physics, University of Nijmegen,
Nijmegen, \\
The Netherlands}

\begin{document}

\maketitle
\begin{abstract}
A review is given of the various Nijmegen potentials. Special attention
is given to some of the newest developments, such as the extended soft-core
model, the high-quality potentials, and the Nijmegen optical potentials for
NN.
\end{abstract}

\footnotetext{
Invited talk given by J.J. de Swart at the XVth European Conference on Few-Body
Problems in Physics, held at Pe\~niscola, Spain, 5--9 June 1995.}

\section{Introduction}         \label{I}
A large part of the efforts of the Nijmegen group in the last decennia
has been concentrated
on the study of the baryon-baryon~\cite{7704,7103,8312},
as well as the antibaryon-baryon interaction~\cite{8306,LEAP94}.
In first instance this has been the construction of
potentials~\cite{7502,7602,7804,7707,8901},
but later also partial-wave analyses (PWA) of the
experimental scattering data~\cite{11,12} were performed.
The knowledge obtained in
these PWA's was then applied again in the construction of new, improved
potentials~\cite{9305,9304,LEAP94,RT}.
This interplay between potential construction and PWA has
has turned out to be very fruitful.

We considered extensively the
baryon-baryon channels with strangeness \mbox{S = 0, 1, and 2}.
The potentials we constructed for the (non-strange) NN-channels
were in the beginning all ``NN-potentials''.
We say here explicitly ``NN-potentials''.
We mean with this something else then when we say
``$pp$- or $np$-potentials''.
In NN-potentials charge-independence is assumed for the nuclear part of the
potential.
For the exchanged mesons and for the nucleons averaged iso-multiplet
masses are used, such as the average
pion mass $m = (2m_+ + m_0)/3 = 138.4$ MeV,
the nucleon mass $M = (M_p + M_n)/2 = 938.93$ MeV, etc.
The $I = 1$ part of the
NN-potentials was always obtained by fitting the $pp$-data.
Lately we have been constructing $pp$-, as well as $np$-potentials. In such
potentials the mass differences are properly
taken into account. In $pp$-scattering
there is only $\pi^0$-exchange, while in $np$-scattering one must
introduce $\pi^+$- as well as $\pi^0$-exchange.

In our efforts to describe and understand the inelasticity in
$pp$-scattering above the various pion production thresholds, we have studied
also the coupled $I = 1$ NN- and N$\Delta$-channels.

In order to describe the elastic $\Lambda p$ scattering below, as well
as above the $\Sigma$-production threshold, and the elastic and
inelastic $\Sigma p$ scattering, we have been constructing
hyperon-nucleon (YN-) potentials~\cite{7103,7602,8901}.
Charge independence breaking was
partially taken into account by using the correct $\Sigma$-thresholds,
and by introducing explicit $\pi\Lambda\Lambda$-couplings via the
mechanism of $\Lambda\Sigma^0$-mixing. \\
At various times also the $Y = 0$ potentials~\cite{G1,G2,G3} such as
$\Lambda\Lambda$, $\Xi N$, $\Sigma\Sigma$, etc.\ were considered. \\
The baryon-antibaryon ($\overline{\rm B}$B) potentials~\cite{8306,RT,LEAP94}
were constructed to
describe the large amount of elastic $\overline{p} p$-scattering data,
and the quasi-elastic $\bar{p}p \rightarrow \bar{n}n$
charge exchange data. Potentials~\cite{PaulT,I} were also
constructed to describe the various strangeness exchange reactions,
such as $\bar{p}p \rightarrow \bar{\Lambda}\Lambda$, etc. \\

All Nijmegen NN-potentials (except the HQ-potentials) were developed
with in the back of our minds the extension to the YN-channels. This
required treatments which could be generalized to the other channels
with the help of SU(3). One needs therefore to include exchanges
of all mesons of the same meson nonet. Next to the $\pi$-meson
one needs to include the exchanges of $\eta$ and
$\eta^{\prime}$-mesons. Next to the $\rho$- and $\omega$-meson one needs
to take account of the $\phi$-meson.

The various Nijmegen potentials can be grouped into several classes.
These are
\begin{itemize}
\item {\bf Hard core potentials.} \\
   Important examples are the potentials Nijm~D~\cite{7502,7602}
   and Nijm~F~\cite{7804}.
\item {\bf Soft core potentials.} \\
   We think here of Nijm78~\cite{7707} and its update Nijm93~\cite{9305}.
\item {\bf Extended soft core potentials.}
\item {\bf High-Quality $pp$- and $np$-potentials.} \\
   We would like to call a potential a HQ-potential, when compared
   with the experimental data it has a $\chi^2/N_d < 1.05$. The only
   examples~\cite{9305,9304} of this class are Nijm~I, Nijm~II, and Reid93.
\item {\bf Optical potentials.}
\end{itemize}

\noindent
We would like to point out here the existence of the NN-OnLine facility
of the Nijmegen group~\cite{url}, which is accessible via the World-Wide Web.
There one can obtain the various Nijmegen e-prints, the fortran codes
for some of the Nijmegen potentials, the deuteron parameters and
the deuteron wave functions, the phases obtained from the Nijmegen PWA
and from the Nijmegen potentials, and predictions for many of
the experimental quantities. A direct comparison of these predictions
with the Nijmegen NN-data base is also possible.

\section{Hard Core Potentials}   \label {II}
The Nijmegen~D potential~\cite{7502} was one of our first hard core
potentials that had an acceptable $\chi^2$ with respect to the
experimental NN-data. The extension~\cite{7602} to the YN-channels
did describe the YN-data well. Because of the succes of this potential
in hypernuclear physics~\cite{Bando}
we used this potential also for the construction
of $\overline{\rm B}$B-potentials~\cite{8306,PaulT}.
The elastic and charge exchange
$\overline{p} p$-scattering data are well described in the Nijmegen
coupled-channel model~\cite{8306}. Our wish to study
also the strangeness exchange
reaction $\overline{p} p \longrightarrow \overline{\Lambda} \Lambda$
was the main reason for using the Nijmegen~D model in these anti-particle
reactions, because this Model~D had been tested in the YN-channels.
The Nijmegen soft-core potential~\cite{7707} was then already
available, but was not tested yet in YN.\\

It is very interesting to look at some of parameters of this, now 20
years old, Nijm~D potential. The $\pi$NN-coupling constant was determined by
fitting to the experimental NN-scattering data from before 1969, using
the PWA of the Livermore group~\cite{Arndt}.
We obtained $f^2/4\pi = 0.074$. This must be compared with the present
best value  $f^2/4\pi = 0.0748$. This 20 year old result shows, that
the recently obtained low value for the $\pi$NN-coupling constant is
not due to recent experimental data, but results from our more
sophisticated handling of the data. It was unfortunate that at that
time in 1975, when we found this low value,
we were so brainwashed by the $\pi$N-community in thinking
that $f^2/4\pi \simeq 0.080$, that we did not take this result very
seriously. It lasted till 1984 before we were convinced that the
$\pi$NN-coupling constant was indeed so low~\cite{Karls}.\\
This low value for the $\pi$NN-coupling constant resulted in a very
good deuteron in Nijm~D. For the $d/s$ ratio we then found $\eta = 0.0251$.
This must be compared with the present best value $\eta = 0.0252(1)$.\\
The value $\rho (- \varepsilon , -\varepsilon ) = 1.776$ fm for the effective
range at the deuteron pole must be compared with the present value
$\rho ( - \varepsilon , - \varepsilon ) = 1.764(3)$ fm.\\
Other consequences of this all are: the pretty good values for the
$d$-state probability $p_d = 5.9$\%, and the uncorrected value for the
electric quadrupole moment $Q_0 = 0.272$ fm$^2$. These are in good
agreement with the most recent guesses $p_d = 5.7(1)$\% and
$Q_0 = 0.271(1)$ fm$^2$.\\
Another Nijmegen hard core potential that is worth looking at, is
the Nijm~F model. This model has an NN as well as an YN-version.
In fact even an YY-version~\cite{G2} was constructed, but unfortunately never
published. Recent calculations~\cite{Motoba} show,
that the YN-version of Model~F
reproduces many of the features in hypernuclear physics.

\section{Soft Core Potentials}   \label {III}
The Nijmegen soft-core OBE-potential Nijm78~\cite{7707} and its updated
version Nijm93~\cite{9305} are based upon Regge pole theory. The
corresponding YN-version~\cite{8901} was published in 1989.
This potential has also been applied to antibaryon-baryon
scattering~\cite{12,RT,I}, where very good descriptions of the
various reactions, such as elastic scattering, charge exchange
scattering, and strangeness exchange scattering, have been obtained.\\
One of the attractive features of this potential is that the
coordinate space version and the momentum space version are
exactly equivalent~\cite{9405}. This at the cost of having only a minimal form
of non-locality. In the triton this minimal non-locality has a
100 keV effect on the binding energy~\cite{9304}.

\section{The Extended Soft Core Model} \label{VII}

We next mention an important improvement on the soft-core OBE-model.
Inspired by the chiral quark model, see for example~\cite{Wei68},
and duality~\cite{7704,Dol68,Swa89},
recently there has been constructed the extended soft-core model
(ESC model) for the NN interaction.
The first results with this ESC model were reported
in Refs.~\cite{Rij93,Sto94}.
The ESC-model contains, besides soft OBE potentials of~\cite{7707},
also contributions from two-meson exchange diagrams
($\pi\pi,\pi\rho,\pi\varepsilon$, etc.)~\cite{Rij91,Rij95a},
and from one-pair and two-pair diagrams. The latter are generated
through pair-vertices
($\pi\pi,\pi\rho,\pi\varepsilon$ etc.)~\cite{Rij95b}.
These meson-pair vertices are,
except for a few, all fixed by heavy meson saturation. This way an excellent
fit to the NN single energy PWA is achieved with a restricted
set of free parameters. This model is still under construction. A
preliminary fit is reached with $\chi^{2}_{p.d.p.} = 1.08$.

\section{High Quality Potentials}   \label {IV}
In the various Nijmegen partial wave analyses~\cite{11} of the NN-scattering
data we can describe these data with a $\chi^2/N_d \simeq 1.0$.
This means that with a potential model description this will also
be about the limit. A measure for the quality of potentials is
therefore the difference with 1.0 for the value of $\chi^2/N_d$.
Let us define high quality (HQ) potentials as having
$\chi^2/N_d < 1.05$ when compared directly with the experimental data.
We constructed in Nijmegen three potentials of this HQ-type~\cite{9305,9304}.
These are the potentials Nijm~I, Nijm~II, and Reid93. They have
all the excellent value $\chi^2/N_d = 1.03$. \\

\noindent {\bf Nijm~I} \\
The Nijmegen soft-core Nijmegen potential Nijm78 has been the starting
point in the construction of the high quality NN-potentials. This
soft-core potential gives already a reasonable good description of all
partial waves. In each partial wave separately the description can
be improved and made excellently, when we allow the parameters of the potential
to be adjusted in each partial wave separately. This leads to the
Nijm~I potential. This potential has a minimal form of non-locality
in the central potential only. There
\begin{displaymath}
     V_c (r) = V(r) - \frac{1}{2m} (\bigtriangleup V' + V'\bigtriangleup)
\end{displaymath}

\noindent {\bf Nijm~II and Reid93} \\
The Nijm~II potential is similar to the Nijm~I potential, but with
all non-locality in each partial wave removed, i.e.\ $V' \equiv
0$. The Reid93 potential
is an update of the old Reid potential (see ref~\cite{9305}).
The singularities in this
potential at the origin are removed by introducing formfactors.
The main difference between the Nijm~I and the Reid93 potential is
just these formfactors. In the Nijmegen potentials an
exponential form factor has been used
\begin{displaymath}
     F(k^{2}) = \exp(-k^{2}/\Lambda^{2}),
\end{displaymath}
while in the Reid93 potential we used
\begin{displaymath}
     F(k^{2}) = (\Lambda^{2}/(\Lambda^{2} + k^{2}))^{2}\ .
\end{displaymath}

The presence of form factors is due to the spatial extension of the nucleons
and the mesons. Quarkmodels using harmonic oscillator interactions
between the quarks will lead quite naturally to exponential form
factors~\cite{Fe}.

\section{The Deuteron} \label {V}

\begin{figure}[b]
\vspace{4.5cm}
\includegraphics{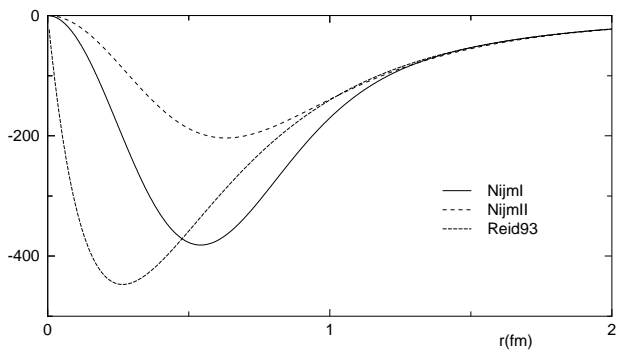}
\caption{The tensor potential connecting the $^3S_1$ and
$^3D_1$ partial waves.}
\label{fig1}
\end{figure}

\begin{figure}
\vspace{8cm}
\includegraphics{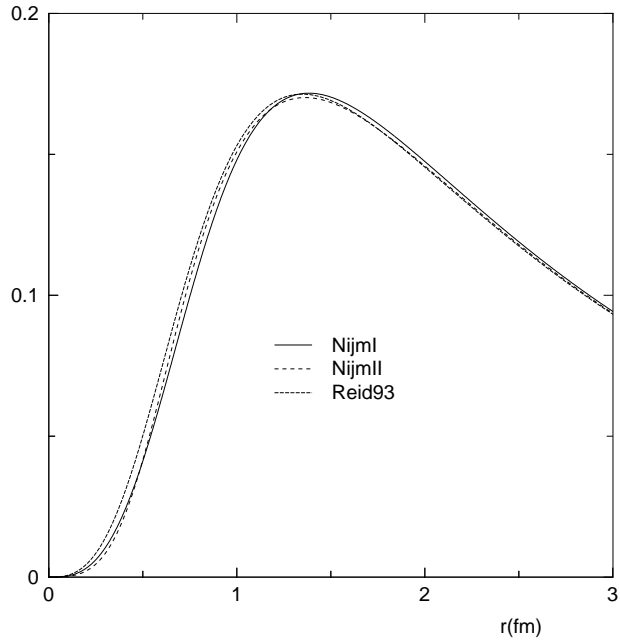}
\caption{The deuteron $d$-state wave function $w(r)$.}
\label{fig2}
\end{figure}

\begin{figure}
\vspace{8.5cm}
\includegraphics{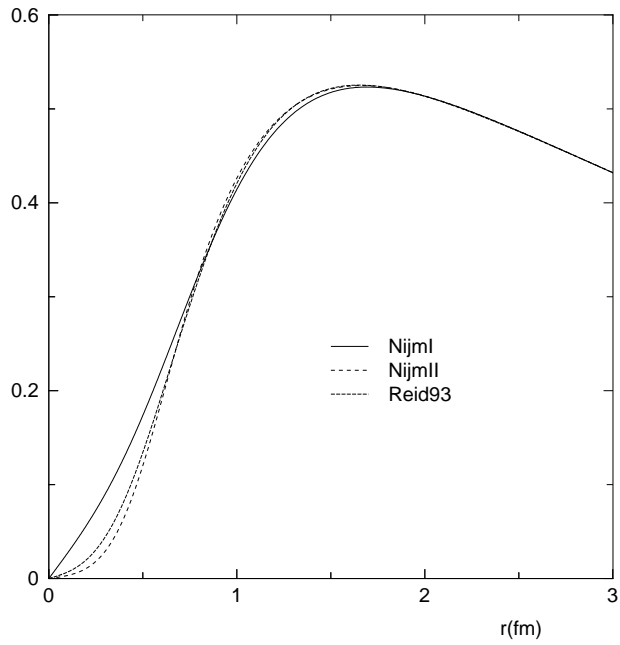}
\caption{The deuteron $s$-state wave function $u(r)$.}
\label{fig3}
\end{figure}

It is instructive to compare the tensor potentials of the 3
HQ-potentials. In doing this we must keep in mind that the quality
of the description of the NN-data is in all three cases essentially
the same.
In Fig.~\ref{fig1} we plot the tensor potential connecting the $^3S_1$ and
$^3D_1$ partial waves. We see that in the inner region, $r < 1$ fm, these
potentials are quite different. In Fig.~\ref{fig2} we plot the deuteron
$d$-state wave function $w(r)$. It is remarkable that these
wave functions are essentially the same, despite the fact that the
tensor potentials are different.
In Fig.~\ref{fig3} we plot the deuteron $s$-state wave function
$u(r)$. We see again the agreement between these wave functions
for large values of $r$ but for values of $r < 0.6$ fm we spot some
interesting differences.

There is a fantastic agreement between the values of the deuteron parameters
as determined in our PWA's and the values for the same parameters
as given by the HQ-potentials. These deuteron parameters
are~\cite{Te}:\linebreak[4]
the $d/s$ ratio $\eta = 0.0252(1)$ and \\
the effective range at the
deuteron pole $\rho(-\varepsilon,-\varepsilon)=1.764(3)$ fm. \\
These values follow directly from the scattering data, and can
be considered as the best experimental determinations. More surprising is,
that the value of the $d$-state probability $p_d = 5.7(1)$\%
and the value of the uncorrected electric quadrupole moment
$Q_0 = 0.271(1)$ fm$^{2}$ are more or less unique. This value of
$Q_0$ does not agree at all with the experimental value~\cite{Bi}
$Q=0.2859(3)$ fm$^2$.
The difference $Q-Q_0 = 0.015$ fm$^2$ needs to be explained in terms of meson
exchange currents, relativistic effects, contributions of
$Q^6$-states, etc.

\section{Optical Potentials} \label {VI}

Below the threshold(s) for pion-production the NN-potentials
are real. When one wants to describe the inelasticities in the
scattering above these thresholds, then one has to go to either a
complicated coupled channel description or one has to introduce
an optical NN-potential: $V=V_R - i V_I$.

The influence and the importance of the imaginary part of the optical
NN-potential can clearly be seen in our PWA~\cite{Rene}
of the $np$-scattering data below
$T_{L}=500$ MeV. In Table~\ref{table1} we give for various energy
intervals the value of $\chi^2$ obtained in our PWA, and the
increase in the value of $\chi^2$ when we omit the
imaginary part of the potential.

\begin{table}[htb]
\caption{The values of $\chi^2$ obtained in our PWA of the $np$ data
  for various energy intervals. Also given are the number of data in that
  interval and the increase $\Delta \chi^{2}$ in the value of $\chi^2$, when
   we omit the imaginary part of the potential.}
\label{table1}
\begin{center}
\begin{tabular}{r@{-}lrrr}
\multicolumn{2}{c}{Interval} & \# data & \multicolumn{1}{c}{$\chi^2$}
                   & $\Delta \chi^2$ \\ \hline
0 & 350   & 2549 & 2519 & 11 \\
350 & 400 &  254 &  306 & 13 \\
400 & 450 &  319 &  337 & 105 \\
450 & 500 &  866 &  905 & 651 \\
0 & 500   & 3988 & 4067 & 880 \\ \hline
\end{tabular}
\end{center}
\end{table}

Looking at this Table~\ref{table1}  we come to the conclusion that
the use of a purely {\bf real} potential works very well up to
$T_L = 400$ MeV, it works well up to $T_L = 450$ MeV, and one can
say that it works reasonably up to $T_L = 500$ MeV.

Let us next make from the real HQ-potentials optical potentials by adding
to them the same imaginary part as was used in our PWA of the $np$-data
below $T_L = 500$ MeV.
When we compare now with the experimental $np$-data below 500 MeV we find
for the Nijm~I and Nijm~II optical potentials $\chi^2/$\# data $\simeq 2.4$
to 2.45.
This is quite a lot larger than the minimum $\chi^2$-values obtained in
our PWA of these data. It turns out that especially the
$^1F_3$ and $^1D_2$ partial waves are responsible for the big rise
in $\chi^2$.

\begin{figure}
\vspace{4.5cm}
\includegraphics{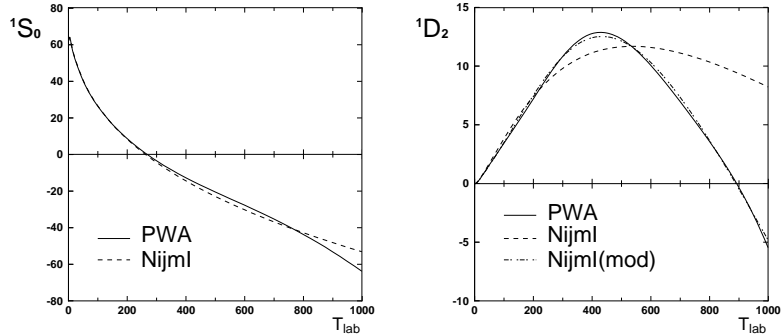}
\caption{$^1S_0$ and $^1D_2$ phase shifts for the optical Nijm~I
  potential and for the modified version.}
\label{fig4}
\end{figure}

We are looking into the possibility to construct optical potentials that are
good up to 1 GeV. In Fig.~\ref{fig4} we give the
$^1S_0$ and the $^1D_2$-phase shifts as determined in a preliminary
PWA of all $np$-data below 1 GeV. In the same figures are also
plotted the prediction of the optical Nijm~I potential. For the
$^1S_0$-phase the description is pretty good up to 1 GeV.
For the $^1D_2$-wave one notices quite large differences. However,
after refitting we get the modified Nijm~I optical potential
(Nijm~I(mod)), which gives
a very good fit to the $^1D_2$-phase shift up to 1 GeV.
We think that it will be not too difficult to produce an
optical potential which fit the $np$-data up to 1 GeV.

\begin{acknowledge}
Part of this work was included in the research program of the Stichting
voor Fundamenteel Onderzoek der Materie (F.O.M.) with financial
support from the Nederlandse Organisatie voor Zuiver-Wetenschappelijk
Onderzoek (N.W.O.).
\end{acknowledge}


\SaveFinalPage
\end{document}